\documentclass[sigconf]{acmart}
\usepackage{algorithm}
\usepackage{algpseudocode}
\usepackage{caption}
\usepackage{subcaption}
\usepackage{ragged2e}
\usepackage{tabularx}
\usepackage[shortlabels]{enumitem}

\newcolumntype{L}[1]{>{\hsize=#1\hsize\RaggedRight} X}

\AtBeginDocument{%
  }

\copyrightyear{2024}
\acmYear{2024}
\setcopyright{acmlicensed}\acmConference[CHI '24]{Proceedings of the CHI Conference on Human Factors in Computing Systems}{May 11--16, 2024}{Honolulu, HI, USA}
\acmBooktitle{Proceedings of the CHI Conference on Human Factors in Computing Systems (CHI '24), May 11--16, 2024, Honolulu, HI, USA}
\acmDOI{10.1145/3613904.3642074}
\acmISBN{979-8-4007-0330-0/24/05}

\usepackage{xcolor}
\newcommand{\rev}[1]{#1}

\newcommand{\egboxwide}[1]{
\smallskip
\noindent
\fbox{
        \parbox{0.97\linewidth}{
        \vspace{0.5ex}#1\vspace{0.5ex}}
      }
\smallskip
}

\makeatletter
\newcommand\footnoteref[1]{\protected@xdef\@thefnmark{\ref{#1}}\@footnotemark}
\makeatother

\begin{document}

\title{Automatic Macro Mining from Interaction Traces at Scale}

\author{Forrest Huang}
\email{forresthuang@google.com}
\affiliation{%
  \institution{Google Research}
  \country{USA}
}
\author{Gang Li}
\email{leebird@google.com}
\affiliation{%
  \institution{Google Research}
  \country{USA}
}
\author{Tao Li}
\email{tlinlp@google.com}
\affiliation{%
  \institution{Google Research}
  \country{USA}
}
\author{Yang Li}
\email{liyang@google.com}
\affiliation{%
  \institution{Google Research}
  \country{USA}
}

\renewcommand{\shortauthors}{Huang et al.}

\begin{abstract}
Macros are building block tasks of our everyday smartphone activity (e.g., "login", or "booking a flight"). Effectively extracting macros is important for understanding mobile interaction and enabling task automation. These macros are however difficult to extract at scale as they can be comprised of multiple steps yet hidden within programmatic components of mobile apps. In this paper, we introduce a novel approach based on Large Language Models (LLMs) to automatically extract semantically meaningful macros from both random and user-curated mobile interaction traces. The macros produced by our approach are automatically tagged with natural language descriptions and are fully executable. We conduct multiple studies to validate the quality of extracted macros, including user evaluation, comparative analysis against human-curated tasks, and automatic execution of these macros. These experiments and analyses demonstrate the effectiveness of our approach and the usefulness of extracted macros in various downstream applications.
\end{abstract}



\begin{teaserfigure}
    \includegraphics[width=\textwidth]{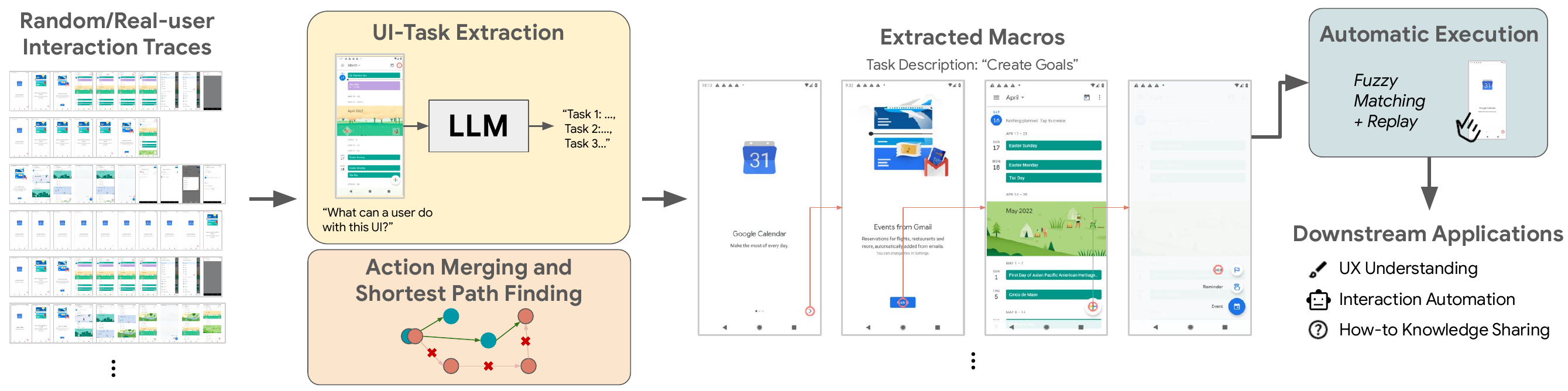}
    \caption{We propose a novel system that can automatically extract semantically meaningful and replayable macros that reflect useful tasks on apps from random or human-curated interaction traces. Colored components in this figure represents the primary technical contributions of this paper.}
    \Description{This is a teaser figure of the paper that shows the overall workflow of our system. On the left, a number of random or real-user interaction traces of mobile apps is shown, with many sequential screenshots pictured. These traces have an arrow pointing towards the second component of the figure, which is divided into the top part that represents UI-Task Extraction, and the bottom part that represents Action Merging and Shortest Path Finding. The top part depicts a screenshot being fed into an LLM, and the LLM outputs several tasks. The bottom part depicts a graph with some edges removed when a new graph is merged with the old one. These components point to the third part of the figure, which depicts an extracted macro, along with a few screenshots of mobile apps. The screenshots are from the calendar app of a mobile app, and each screenshot is a result of clicking on the previous button. The task description is “create goals”, and these screenshots depict four steps of how a user might create a goal in the app’s UI. Finally, this extracted macro gets fed into the Automatic execution system depicted in the right side of the figure, which performs fuzzy matching and replays the macro on a real device. This workflow allows for various downstream applications such as interactive task understanding, interaction automation and how-to knowledge sharing.}
 \end{teaserfigure}
\maketitle

\section{Introduction}
The interaction between users and mobile apps can be abstracted, understood, and studied at many levels. At the lowest level, for example, users' granular motor movement on mobile devices can be studied and characterized by the important Fitts' Law and its numerous derivatives introduced by the HCI community. On the other end of this spectrum, mobile apps can be organized by developer-defined Views (iOS) or Activities/Intents (Android) that group app code and UIs with similar programmatic logic and functionalities. 

An important abstraction among this spectrum is the notion of \emph{macros}. Macros represent well-encapsulated units of users' engagement with apps with certain needs and/or in certain contexts. For example, a macro for a user in a to-do list app might be `adding a to-do list item'. This macro represents a specific user need and covers how the app might address this user's problem with a collection of actions and parameters in the app. Completing this macro in an app would require the user to perform multiple clicks to reach multiple screens and set multiple values/parameters (e.g., the due date of the item) on each of them. Because macros are important constituents of our everyday smartphone activities, extracting them is highly valuable for a number of purposes, including interaction automation, how-to knowledge sharing~\citep{coscriptor}, as well as understanding interactive tasks.

Although extensive work has been conducted in creating macros using techniques such as Programming by Demonstration (PBD)~\cite{conversational-crowd, pumice, sugilite}, little progress has been made in automatically extracting high-level and functionally complex macros from interaction traces. Mining these traces, which are often abundant from crawling, is a crucial step towards extracting macros at scale. However, automatically extracting macros that represent meaningful user tasks from these traces is unfortunately not straightforward. The mappings between macros to programmatic invocations are often non-linear and non-consistent---a macro can involve multiple function calls, spanning across multiple views, yet some views can support multiple macros. It is therefore difficult to directly instrument or reverse-engineer these macros based on app source code or user recordings which were employed in prior work~\cite{shirazi,alharbi}. Additionally, many of these macros are context-dependent. As a result, it is difficult to derive a unified taxonomy of macros, compared to more granular classifications such as classes of icons or UI elements~\cite{liu, semantic-elem, enrico}, which makes global aggregations across the dataset difficult. While there are existing efforts that create macros via crowdsourcing~\cite{motif}, it remains out of reach for capturing a large proportion of functionalities of each app and handling a vast variety of apps.

Recent advances in Large Language Models (LLMs) have enabled a new class of methods and models that can understand and interact with mobile apps with an unprecedented level of intelligence. LLMs have repeatedly been shown to possess common knowledge about mobile UIs and users' daily tasks, enabling applications such as conversational interactions, screen summarization for accessibility, and multi-step task grounding and planning~\cite{bryan}. Motivated by these exciting findings, we investigate using LLMs to automatically extract macros from mobile apps. By `inspecting' UIs converted to an HTML format, similar to prior work, and prompting LLMs appropriately, we show that LLMs can effectively extract semantically meaningful macros from apps that cover many interactions by describing these macros flexibly in natural language. To make these macros automatically executable, we take further steps by feeding them back into the LLM to identify elements on the screen and parameters required to fulfill the macros, and synthesizing multiple execution traces to distill optimal click paths to execute the macros. We experiment with our approach on three datasets: RICO~\cite{rico}, MoTIF~\cite{motif}, and a dataset of random crawls of apps that we created. We conduct a user study to evaluate the quality of extracted macros; we compare extracted macros against human-curated tasks quantitatively; and we test the automatic-executability of these macros in a live environment. These experiments and analyses show that our approach is effective in extracting meaningful macros from arbitrary interaction traces. Our main contribution is three-fold:

\begin{itemize}
    \item We contribute a novel approach of using LLMs with a trace-based chain-of-thought technique and optimal path synthesis to effectively extract large-scale, meaningful \emph{macros} from interaction traces that are abundant in existing datasets.
    \item The macros we extracted enrich existing mobile datasets. In particular, we extracted a large dataset of 23,777 macros from RICO~\cite{rico}, an existing large-scale UI dataset that is widely used for mobile interaction analysis and modeling. This new dataset of extracted macros will be publicly released\footnote{\url{https://github.com/google-research/google-research/tree/master/macro_mining}}.
    \item Beyond evaluating our extracted macros with human users and against existing large-scale datasets quantitatively, we executed our extracted macros in live environments. These evaluation results provide evidences for high efficacy of the extracted macros in supporting downstream applications in realistic scenarios. 
\end{itemize}

\section{Definition of Macros and Prior Work}
To fully understand the value of \emph{Macros}, and consequently automatically extracting and executing them, we first define macros used in this work in the context of user-app interactions. The interaction between a user and a mobile app can be decomposed into sets of chronologically ordered actions $a_i$, screens that the actions was initiated from $s_i$, and the UI elements $e_i$ that were involved in the actions. In each interaction session, a user performs an action on each screen on a specific element or a set of elements, leading to the next screen when this process repeats again. Each instance/session of interaction can thus be viewed as a \emph{trace}, where the trace consists of a sequence of triplets of $n$ actions, screens and elements $(a, s, e)_{1...n}$. 

\rev{To analyze and aggregate collections of multiple traces, prior research have considered these components as various \emph{units of interaction}. A significiant amount of prior research have investigated UI elements $e_i$ as independent units by annotating and classifying them~\cite{liu,semantic-elem}.
Researchers have also grouped multiple views and elements by their corresponding programmatic implementations~\cite{alharbi}.}

This work focuses on a more user-centric definition for units of interaction---\emph{Macros}. A macro could be defined as a series of actions $a_i$ that collectively performs a semantic task or achieves a user goal. This means a macro could be described in natural language, corresponds to a set of actions in a sequence, and all actions serve the sole purpose of completing a task meaningful to the user (e.g., booking a flight, checking the wait time at a restaurant). The collections of views and actions in a macro are often orthogonal to those arranged by developer-defined abstractions; an activity in Android, for instance, can support multiple macros, yet each macro can span across multiple activities. \rev{Automatically extracting them also requires a different type of computational understanding, which includes holistic knowledge about the interaction between user tasks, usage context, and UI components. We summarize the differences between our work and the aforementioned related work in Table~\ref{tab:comp}.}

\begin{table*}[h!]
  \caption{\rev{Summary of differences in interaction units and requirements for extracting them between our work and prior work.}}
  \Description{This table summarizes the differences between our work and prior work in terms of interaction units and requirements for extraction. Our work studies multiple UIs grouped by user tasks and use-cases, and requires a holistic understanding of UIs, tasks and contexts in users' app usage.}
  \label{tab:comp}
  \begin{tabular}{p{3cm}p{5cm}p{6cm}}
    \toprule
    Research Work & Studied Interaction Units & Requirements for Extraction \\
    \midrule
    Semantic \newline Classification \newline \cite{liu, clay, semantic-elem, enrico} & Individual UI elements and screens & Understanding correspondence between graphical, geometric and textual properties of UI elements and a fixed set of semantic UI concepts \\
    \midrule
    UI Embeddings \newline \cite{screen2vec, uibert, actionbert} & Individual and pairs of UI screens & Understanding UI element and screen functionality in context of nearby UIs in interaction \\
    \midrule
    Developer-defined \newline Abstractions~\cite{shirazi, alharbi, app-usage} & UI Activities, Views, and Packages with multiple elements and screens grouped and implemented by developers & Access to source-code or reverse engineering \\
    \midrule
    \textbf{Macro Mining (Ours)} & \textbf{Multiple UIs grouped by \emph{user} tasks and use-cases} & \textbf{Holistic understanding of UIs, tasks, and contexts in users' app usage} \\
  \bottomrule
\end{tabular}
\end{table*}


In the remainder of this section, we compare and contrast various prior work against our proposed approaches. We also review prior attempts of extracting macros and potential applications that macros could enable.

\subsection{Semantic Understanding of Individual UI Elements and Screens}
\rev{Individual UI elements and screens are extensively studied by prior work as interaction units in traces. Prior works have developed two main categories of methods to understand the semantics of these elements in relation to users' interaction intents and needs.} The first category of works classified UI elements~\cite{liu, clay, semantic-elem} and screens~\cite{enrico} into fixed sets of researcher-curated semantic and functional concepts (e.g., text input, `add' icon) with crowd-sourcing and machine-learning models. The second category of works explored representing UI screens and elements through free-form text annotations or ML-model-embeddings in various UI-based learning tasks. Widget Captioning~\cite{widget-captioning} and screen2words~\cite{screen2words} respectively collect human-provided natural language annotations of UI elements and screens. Screen2vec~\cite{screen2vec}, UIBert~\cite{uibert}, and ActionBert~\cite{actionbert} explore the development of embeddings for both UI components and screens through training ML models on prediction tasks of the context and functionality of the interaction units concerned.





\subsection{Developer-defined Abstractions of UIs}
Apps UIs can also be analyzed from the alternative perspective of developer-defined abstractions. In Android apps, developers define Activities, Intents, Services and Layouts, which can provide meaning to various parts of the apps~\cite{shirazi}. These abstractions allow prior work to more easily instrument and analyze them by decompiling the app packages (apks) statically~\cite{shirazi}. These abstractions have also supported the discovery and analysis of design patterns~\cite{alharbi} and task usage patterns~\cite{app-usage} within and across boundaries of apps. Beyond UI-related applications, these abstractions have supported security-~\cite{security} and accessibility-related~\cite{access} applications. Nevertheless, these abstractions can sometimes be misaligned with actual task-based usages of the apps in our use-cases.

\subsection{Defining Macros through demonstration, End-User programming, and Task-based Applications}
Macro is a familiar concept in research works~\cite{pumice, pbd1} and commercial applications~\cite{ifttt} in the area of Programming By Demonstration (PBD), enabling users to record macros to automate complex tasks. Prior works have also coupled the concept of macro-authoring with conversational interaction with automatic~\cite{sugilite} and crowd-based systems~\cite{conversational-crowd}. 

Researchers have also explored the possibility to infer macros directly from natural language \emph{without} user-demonstrations. MoTIF introduces a crowd-sourced dataset that contains interaction traces with their task descriptions, and trains model to complete these \emph{grounding} tasks of performing correct actions on the UI screens from text descriptions~\cite{motif}. SAVANT is a system that automatically matches user-provided task descriptions with relevant app UIs, providing shortcuts for users to perform tasks hidden within mobile apps~\cite{savant}. While these models and applications are highly relevant to some of the applications we envision our system can support, none of these prior work \emph{comprehensively discover and generate multiple possible tasks (macros)} at a large-scale from merely observing interaction traces. 



\subsection{LLM-based UI Applications}
Large Language Models (LLMs) have demonstrated impressive general reasoning ability~\cite{gpt4}, and this ability was shown to be extensible and applicable to the UI domain, with researchers using LLMs to enable many types of conversational UI-based interactions without fine-tuning, including question-answering, summarization, and grounding on mobile UIs~\cite{bryan}. Spotlight~\cite{spotlight} and Pix2struct~\cite{pix2struct} have further shown that a single pre-trained model can achieve state-of-the-art performance in multiple downstream UI tasks. These works have provided evidence for LLMs being able to operate in the UI domain when provided with reasonable representations (e.g., a simplified HTML representation introduced in~\cite{bryan}), which inspires our approach of using LLMs to effectively extract semantic tasks from UIs.

\section{Macro Extraction and Execution}
\label{sec:pipeline}
Our extraction system consists of two major steps: \emph{task extraction} and \emph{action merging}. We first utilize recent advances in LLMs and perform chain-of-thought reasoning~\cite{aichains, cot} at each screen in the UI interaction traces. This step further contains three sub-steps:

\rev{
\begin{enumerate}[(a)]
    \item \emph{Task Discovery}: LLM discovers user-tasks on each UI screen;
    \item \emph{Action Grounding}: LLM predicts the relevant action for completing the user-tasks, and;
    \item \emph{Parameter-finding}: LLM predicts additional info required for the user-tasks.
\end{enumerate}
}

\begin{figure*}[h!]
  \centering
  \includegraphics[width=\linewidth]{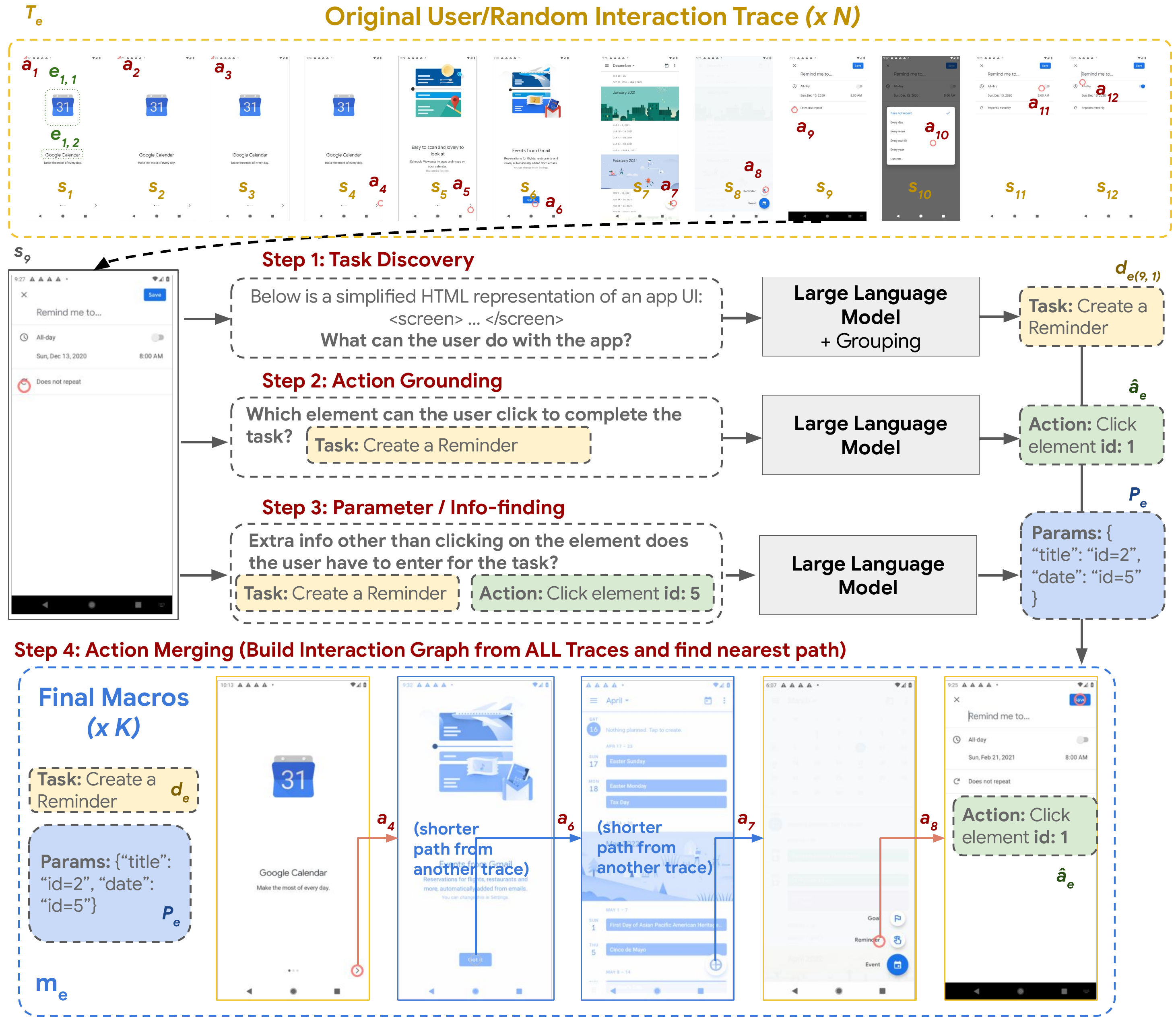}
  \caption{Full macro extraction system for a single example macro: `Create a reminder'. Our system extracts K macros from all N traces for each app in the actual datasets.}
  \Description{This is a large figure that demonstrates how a single example macro: `create a reminder', was extracted. At the top of the figure, there is a row of screenshots which represents an original user or random interaction trace as a sequence of screenshots. We extend an arrow from one of the screenshots to a larger version of it in the second row, in which this screenshot is indicated as an input to three steps for extracting the macro. The first step is task discovery, where we ask the LLM to discover tasks to be done on the app UI by asking it ‘what can the user do with the app?’ This is a box in the figure and it is pointed towards the large language model box, which outputs a task of `create a reminder'. The second step, which is a second box in the figure, asks the LLM which element can the user click to complete the task, along with the task of ‘create a reminder.’. The LLM then responds with the action to: “Click on the element with an id of 5”. Then, in step 3 the LLM also generates the parameter to complete the macro based on the task and action. The final step of our system builds an interaction graph from all traces in this app, and then combine the trace with the shortest path with the extracted task, parameter and actions, which is shown in the bottom part of the figure.}
  \label{fig:pipeline}
\end{figure*}

\rev{
The second major step of the system merges these action traces using their screen and action data to form per-app interaction graphs, and consequently generates an optimized final set of macros based on shortest paths on the graphs. 

This entire extraction process is illustrated in Figure~\ref{fig:pipeline}. We walk through this process of extracting an example macro `Create a reminder' (shown in Figure~\ref{fig:pipeline}) in each of the following subsections while explaining the implementation details of the system. Finally, we also contribute a Macro Replayer that can flexibly execute the extracted macros. 
}

\subsection{Notation, Extraction Inputs, and Extraction Outputs}
\rev{
The input to our extraction system is a collection of user interaction traces. Each trace $T$ comprises of multiple pairs of screens $s_t$ and actions $a_t$ that leads to the next screen $s_{t+1} = a_t(s_t)$. Each screen $s$ has UI elements $e_{t, i}$. These traces can either be user-provided or automatically-crawled. For example, the trace $T_e$ in Figure~\ref{fig:pipeline} contains a series of screens from the \emph{calendar} app UI, with actions performed by a random automatic crawler. $s_1$ in this trace is the landing page of the calendar, which contains multiple UI elements including the calendar icon $e_{1, 1}$. 

$T_e$ contains a few random actions $a_1$, $a_2$, and $a_3$ that were performed on the first screen (such as pressing system buttons or scrolling vertically\footnote{Scrolling is handled by starting from an element and scrolling down by a fixed amount.} on the page). However, these actions do not have any effects on the UI screen, hence the UI screens from $s_1$ to $s_4$ are identical. On $s_4$, the crawler performed a click action $a_4$ on the `arrow' button on the bottom right, advancing to the next screen of the calendar onboarding page as $s_5$. On $s_5$, $a_5$ was another click on the right arrow button, resulting in the final onboarding page $s_6$. On $s_6$, the `Got it' button was clicked ($a_6$), and the UI advances to the main calendar page $s_7$. This cycle continues until the crawler reaches a pre-determined maximum number of 30 actions, or if an action leads to the crawler exiting the app. Alternatively, a trace can be taken directly from a dataset in the same format, such as from RICO~\cite{rico}.

From all traces of an app, our system automatically extracts a collection of macros $M = \{m = (d, \{a\}, \{p\})\}$ where $\{a\}$ is a collection of actions that a user would perform to complete a macro task, $\{p\}$ is the set of additional information required for the task, and $d$ is a natural language description of the macro. In the example, the extracted macro $m_e = (d, \{a_e\}, \{p_e\})$ has the natural language description $d_e$ "Create a reminder", a set of actions $\{a_e\}$ that includes clicking on the `next' button ($a_4$, $a_5$), clicking on the `Got it' button ($a_6$), clicking on the `add' FAB button ($a_7$), clicking on the `Reminder' button ($a_8$), and clicking on the `Save' button (predicted by LLM). This macro also includes the parameters $p_e$ of both the title and date as required information from the user, as well as their corresponding UI elements to enter the information into.
}

\subsection{Task Extraction}
\label{sec:task_extraction}

\rev{
\subsubsection{Task Discovery}
The first step towards extracting macros from traces is observing relevant user-tasks from each of the UIs in the app in the traces. We convert each screen $s_t$ into an HTML representation following~\cite{bryan}\footnote{Each interactive element $e$ is represented by an HTML element, with its position represented as strings that correspond to one of nine grids on the screen (e.g., top left, top right, etc.). The text content or content description of the element is taken directly as the content of the HTML element, and the semantic class of the element~\cite{clay} is converted to the tag of the HTML element.}. We then pass this HTML representation to the LLM to prompt for potential user tasks (Step 1 in Figure~\ref{fig:pipeline}). Taking $s_9$ from Figure~\ref{fig:pipeline} as an example, this is a UI for creating a calendar reminder. The screen is converted into the HTML format that describes all elements on the screen, and was passed to the LLM\footnote{The HTML representation is redacted in the prompt example for brevity, the full LLM inputs/outputs are available in Appendix~\ref{appx:llm}.} for processing and generation:


\egboxwide{
  \begin{minipage}{.97\textwidth}
  \textbf{Prompt 1:} \emph{Below is a simplified HTML representation of a \\ mobile app: \\ <screen> \\ <img id="0" class="cancel image" pos="top left"></img> \\
<button id="1" class="save" pos="top right">save</button> \\
<input id="2" class="title edit" pos="top">remind me to</input> \\ <p id="3" class="first line" pos="top">all day</p> \\
<img id="4" class="tile icon" pos="top left"></img> \\
<p id="5" class="first line" pos="top">sun dec 13 2020</p> \\ ... \\ </screen> \\ What can a user do with the prompt? \\ The user can: -}
    \end{minipage}
  }

The responses from the LLM take the form of `<(task 1)> -<(task 2)> ...', which we can then parse back into a collection of candidate descriptions for macros $d_{t, 1...i}$. In this case, the LLM responds with:

\egboxwide{
  \begin{minipage}{.97\textwidth}
  \textbf{Response 1:}  \emph{Create a reminder \\
  - Edit the reminder title \\ 
  - ...}
    \end{minipage}
  }

We parse the corresponding first description $d_{9, 1}$ as our example macro task description $d_e = $ `Create a reminder', from the first line of the LLM response. We then pair each extracted macro description with \emph{all actions} from the trace that lead the UI from the starting screen of the app to the current screen $a_{1...t-1}$. These are then treated as candidate macros.
The example macro candidate $m'_e$ is ($d_e = $`Create a reminder'$, a_{1...8})$, which includes the random scrolls and button clicks that lead to the reminder creation screen. 
}

Once all macro candidates are extracted from all screens from all traces, our system groups and deduplicates highly similar macros by their descriptions (e.g., `create a reminder' should be the same macro as `add a reminder'). We embed each description with Sentence-T5 (a large transformer-based sentence-encoder)~\cite{st5}, and incrementally group similar text descriptions by their cosine similarities\footnote{We add a text description into an existing group if its similarity with this group falls under a certain threshold and the group is most similar to the description.}. We also filter out macro candidates that have generic and/or short descriptions with heuristics defined in Appendix~\ref{appx:macro_filtering}. We then randomly sample one macro candidate from each of these groups as the candidates for next step of processing. 



\subsubsection{Action Grounding}
\label{sec:action}
\rev{To finalize a set of actionable macros, we need to predict the final actions $\hat{a}$ that complete the macros, since they might not be provided by the source traces. We prompt the LLM with the macro description ($d_e$ in the `create a reminder' example) and the HTML representation of the screen that it arrived at after the clicks ($s_9$ from example, Step 2 in Figure~\ref{fig:pipeline}):

\egboxwide{
  \begin{minipage}{.97\textwidth}
  \textbf{Prompt 2:} \emph{Below is a simplified HTML code of a mobile app: \\ <screen> \\ ... \underline{(same as above)} ... \\ </screen> \\ Which element id(s) should the user click on next to accomplish \\ the task \underline{`create a reminder'}? \\ Respond with only the number(s), or ``None'' if the user can \\ already complete the task on the current page.}
    \end{minipage}
  }

This provides a set of elements $E_t = \{e_t\} \parallel \emptyset$ for the corresponding final action $a_t$ to complete the macro. In our `create a reminder' example, the LLM responds with: 

\egboxwide{
  \begin{minipage}{.97\textwidth}
  \textbf{Response 2: }\emph{1}
    \end{minipage}
  }

This id (1) corresponds to the `save' button in the UI, which is then combined into our macro candidate $m'_e$ as the predicted action $\hat{a_e}$, such that the macro candidate is now $m'_e = (d_e = $`Create a reminder', \{$a_{1...8}, \hat{a_e}\})$.




\subsubsection{Parameter-finding}

If a final action is predicted by the LLM (i.e., the LLM doesn't return `None' in Section~\ref{sec:action}), we further prompt the LLM regarding the parameters (extra information) $\{p_e\}$ needed to complete the candidate macro (Step 3 in Figure~\ref{fig:pipeline}). Each parameter contains a text description and the element that the parameter should be entered into. We use the following prompt that includes the text description and the `save' button's id obtained above in our example `create a reminder' macro:

\egboxwide{
  \begin{minipage}{.97\textwidth}
  \textbf{Prompt 3:} \emph{Below is a simplified HTML code of a mobile app: \\ <screen> \\ ... \underline{(same as above)} ... \\ </screen> \\ The user is trying to complete the task \underline{`create a reminder'}. \\ Other than clicking on \underline{the element with id 1}, list the additional \\ information the user needs to enter in the format of (- (info)). \\ Answer ``None'' if no additional information is needed.}
    \end{minipage}
  }

This gives us the text descriptions for the parameters in the form of `- (description 1) - (description 2)...'. For our example, the LLM responds with:

\egboxwide{
  \begin{minipage}{.97\textwidth}
  \textbf{Response 3:} \emph{- (title) \\ - (date)}
    \end{minipage}
  }

We parse this response and obtain `title' and `date' as the parameters needed for the `create a reminder' example macro. We then prompt the LLM again to obtain the relevant UI element id for each parameter. For example, to obtain the element id for the `title' parameter:

\egboxwide{
  \begin{minipage}{.97\textwidth}
  \textbf{Prompt 4:} \emph{Below is a simplified HTML code of a mobile app: \\ <screen> \\ ... \underline{(same as above)} ... \\ </screen> \\ Where can the user enter \underline{title}? Answer with only the element id, \\or ``None'' if no element matches.}
    \end{minipage}
  }

The LLM response will provide us with all information required for each parameters that includes the element reference and the text description. In our example, the LLM responds with:

\egboxwide{
  \begin{minipage}{.97\textwidth}
  \textbf{Response 4:} \emph{2}
    \end{minipage}
  }

The element with id = 2 corresponds to the element where the users enter reminder titles in the creation page. This final response completes the macro candidates $M' = \{(d, \{a_{1...t-1}, \hat{a}\}, \{p_e\})\}$. With our `create a reminder', the final macro candidate is:
\begin{align*}
m_e = (\text{`Create a reminder'}, \{a_{1...8}, \hat{a_e}\}, \\ \{(\text{title}, \text{element 2}), (\text{date}, \text{element 5})\})
\end{align*}

This entire workflow of extracting macros adapts chain-of-thought reasoning for the LLM~\cite{cot}---we first prompt it to reason about meaningful tasks in UIs, we then prompt it to ground the high-level task to act on certain elements on the UIs; finally, based on the actions, we further prompt the LLM to generate relevant parameters, creating complete sets of relevant information for the macros. We believe the LLM's prior knowledge on user tasks have enabled the generation of meaningful macros beyond merely summarizing or synthesizing content of UIs. 
}

\subsection{Action/Trace Merging and Optimal Path-finding}
\label{sec:optim}
The actions within each macro candidate generated by the first step of our system were taken directly from the interaction traces (i.e., all actions that the user/agent performed leading up to a certain screen $a_{1...t-1}$). However, these actions are likely sub-optimal, such that a user might perform multiple tasks in a single trace, or a random computational agent might have to click around multiple screens before reaching a task-based UI. To address this problem, we build interaction graphs from multiple interaction traces of the same app. This discovers shorter paths within each individual trace and/or between multiple traces, allowing the tasks in the macros to be executed optimally (Step 4 in Figure~\ref{fig:pipeline}). \rev{For our example `create a reminder' macro $m_e$, this part of the system optimizes the action set $a_{1...8}$ from the candidate $m'_e$ above, since it doesn't require 9 clicks to `create a reminder'.
}

\subsubsection{Overall Graph Structure}
To build the interaction graph, we first define a root node $n_r$ as the first node that the app lands on. Each node in this graph refers to an action (\emph{not} a screen). Then, for an action $a_t$ taken in the trace and its corresponding element $e_{a, t}$, we find the existing node $n_{a, t}$ by the current node $e_{a, t}$'s id (or create a new one if it doesn't exist), and add a connection between $n_{a, t-1}$ ($n_{a, 0} = n_r$) and $n_{a, t}$. We also create or find the nodes for each actionable element $e_{t,j}$ in screen $s_t$, and connect the corresponding nodes $n_{j, t}$ to $n_{a, t-1}$. \rev{In the example `create a reminder' macro, the root node contains outgoing edges that corresponds to $a_1$ and $a_4$, since the element for $a_4$ (the right arrow in the bottom right) can be found on the first screen $s_1$.

While an interaction graph typically models screens as nodes and actions as edges, we depart from this paradigm and encode actions as nodes and screens as edges. Each node could be thought of as the state reached \emph{after} performing the action $a_t$ that the node is labeled with. And each edge could be represented by the screen that the action at the \emph{target node} of the edge is performed on. The main advantage for such representation is that the extracted macros are robust to changes in the screens---the merging is done between the actions, in which a node is reached as long as the same action was performed. We are also defining a similarity metric at the action level (i.e., element) as opposed to at the screen level, which allows us to flexibly include different levels of context depending on the action being performed. 

\subsubsection{Node Identification and Merging}
We identify and merge nodes by their \emph{id}s.} Each node's id is a combination of the following attributes of the element that the previous action was performed on: resource\_id, text (content), content description, and class (in Android). If both text and content descriptions are empty, we first traverse the view hierarchy downwards to adopt the combination of text and content descriptions of the element's descendants. If no text was found from its descendants, we traverse up the view hierarchy until a node with text or content description was found and adopt those attributes from this ancestor node. This procedure was inspired by the observation that meaningful text annotations that reflect the functionality of UI elements could be found near them in the view hierarchy, hence uniquely identifying the action-elements. \rev{In our example `create a reminder' macro, this allows us to identify the right arrow for $a_4$ to be accessible in $s_1$, such that in the UI, a user is able to click on the next button immediately after landing on the first screen.}

\subsubsection{Optimal Path-finding}
After an interaction graph was built, breadth-first-search is performed for all macros from the root node, and shortest paths from the root replaces the action set for all macro candidates, forming the final set of extracted macros $M = \{(d, A_{\text{opt}}, \{p\})\}$ where $A_{\text{opt}}$ contains the last node in the original action set (corresponds to LLM-predicted $\hat{a_e}$) and the actions of the shortest path from the root to that node. \rev{In our example `create a reminder' macro, this allows us to skip over $s_2$ to $s_5$, such that the final macro only contains $s_1$ (the first landing page of the UI) and $s_6$ (the final screen of the calendar description page that contains `Got it') to proceed with the rest of the macro. Note that this optimization is performed across multiple traces, allowing \emph{all} shortest path to be found for various macros and functionalities in the UI. This produces the final macro for the example in Figure~\ref{fig:pipeline}: 

\begin{align*}
m_e = (\text{`Create a reminder'}, \{a_4, a_6, a_7, a_8, \hat{a_e}\}, \\ \{(\text{title}, \text{element 2}), (\text{date}, \text{element 5})\})
\end{align*}
}

\subsection{Automatic Macro Replayer}
To automatically execute a macro $m_t$, we perform its associated actions from the landing screen of the app, which includes both the optimal actions to the relevant task screen and the LLM-inferred action for task completion. At each time step $a_t \in A_\text{opt}$, we perform fuzzy matching and match elements that is similar to the original referenced elements with a Jaccard similarity metric, based on a set of text-based attributes of the elements.
Moreover, to allow flexibility and error-tolerance in execution, we check for future actions in the macro if a certain action is not matched on a screen, and skip the actions before the matched future actions. We execute these actions on an Android device with adb\footnote{https://developer.android.com/tools/adb}.
This allows us to fully close the loop and complete the task defined in the macro automatically.

\rev{\subsection{LLM and System Implementation Details}

The LLM used in all steps in the system is based on PaLM2-Bison~\cite{palm2}. For all LLM-generated responses, we use top-p decoding and could regenerate responses if none of the previous ranked generations follow our expected format, or contains hallucinated UI element IDs. However, these LLM-based syntax failures are extremely rare in practice (0.0244\% of the macros), and we remove those that contain them. The rest of the macro extraction and execution system was implemented with Python and Apache Beam framework\footnote{https://beam.apache.org/} for efficient distributed computation. The source code of the this system will be publicly released\footnote{\label{note1}\url{https://github.com/google-research/google-research/tree/master/macro_mining}}.
}

\section{Macro Datasets}
\label{sec:dataset}
Using the system described above, we extracted macros from the RICO dataset~\cite{rico} and the Rehearsal dataset. We took 4,189 traces from the RICO dataset that have complete corresponding annotations in the CLAY dataset~\cite{clay}, with one to a few traces per Android app. Each trace contains the UI hierarchy, screenshot, and the user-performed action in each step of the interaction, allowing us to extract the required elements for macros stated above. We extracted 23,117 macros from these traces, with 8.49 macros from each app on average. Each macro contains 3.41 actions on average. Some representative examples of this dataset is shown in Figure~\ref{fig:rico} and are discussed in detail in Section~\ref{sec:qual}, showcasing our system's applicability in a wide range of apps. We open-source these extracted macros from RICO\footnoteref{note1}.

The Rehearsal dataset contains ~162,000 randomly-crawled mobile interaction traces of 10 apps, with ~16,000 traces per Android app, which means it explores each app in greater depth. Similar to RICO, each trace contains the UI hierarchies, screenshots, and randomly invoked actions on the apps for exhaustively exploring the app computationally using emulators. We sample 1,000 traces per app, and extracted 3,389 macros from the Rehearsal dataset (338.9 macros per app on average). Each macro contains 3.4 actions on average. Some representative examples of this dataset is shown in Figure~\ref{fig:rehearsal}, which demonstrates of system's validity in covering many parts of a single app and its effectiveness when applied to a large number of randomly-crawled, non-user traces. 

\section{Evaluation}
We evaluated the extracted macros with multiple methods targeting multiple aspects of the traces. The two main aspects of the macros to be evaluated are the quality of the text descriptions $d$ and the validity of the UI actions $\{a\}$. This ensures that our extracted macros are \emph{reasonable}, \emph{valid}, and \emph{executable}. We evaluated each of the aforementioned aspects with quantitative experiments, qualitative analysis, and user studies. We also designed this set of evaluations with the goal to help us understand the possibility for supporting potential downstream models and/or applications (further discussed in Section~\ref{sec:application}).

\rev{
\subsection{Task Description Quality}
\label{sec:quant}
To evaluate the quality of the text descriptions $d$ and their relevance to realistic user tasks, we first compare the descriptions extracted by our model and crowd-sourced task descriptions from the test set of the MoTIF dataset~\cite{motif}. Each data example in MoTIF contains a task description $d_{\textbf{MoTIF}}$ and an interaction trace $T_{\textbf{MoTIF}}$ extracted from one of 125 apps. We process the interaction trace $T_{\textbf{MoTIF}}$ with our system and extract a set of macros $M$. Then for each text description $d_i$ of each extracted macro, we compute a \textsc{Rouge-L} F-measure score and \textsc{Meteor} score\footnote{Both \textsc{Rouge-L}~\cite{rouge} and \textsc{Meteor}~\cite{meteor} are established metrics in NLP that compute textual matching.} between $d_i \in M$ and $d_{\textbf{MoTIF}}$ and take the highest \textsc{Rouge-L} and \textsc{Meteor} score (most similar pair) as the score for an interaction trace. The high-level intuition is that our system should extract a superset of text descriptions as the ground-truth in MoTIF, yet should have at least one of them covering the task described by the user closely, which is measured by the \textsc{Rouge-L} score. The text descriptions in the generated macros should also be generally similar to the human-authored descriptions without redundant or hallucinated details, which is measured by the \textsc{Meteor} score.

We then report the average scores of this max calculation of each ground-truth across the entire dataset. To show the robustness of our method against the randomness involved during LLM inference, we repeat the same procedure for five times and report the mean and standard deviation. In addition, we show qualitative examples with low \textsc{Rouge-L} scores in the Appendix~\ref{appx:rouge}. We demonstrate that even for some examples with low \textsc{Rouge-L} scores, the text descriptions generated by our system is still reasonable given the interaction traces for these examples.

To provide reference to our method's performance, we also developed two baselines. The first baseline (Element-Text) is implemented by taking text and content descriptions from all UI elements in the MoTIF traces as pseudo descriptions. A comparison against this baseline will show the efficacy of our model in extracting and reasoning useful tasks beyond repeating text content that already exists in the UI. The second baseline (Random-Trace) is implemented by randomly reassigning generated tasks from another trace and comparing them against the ground-truth, which shows our model's ability to generate task descriptions specific to the provided app instead of only generating syntactically correct but meaningless text.

The \textsc{Rouge-L} and \textsc{Meteor} scores for our method and the two baselines are shown in Table~\ref{tab:rouge}. Our system achieves a \textsc{Rouge-L} of 0.47 and \textsc{Meteor} of 0.38 for this evaluation, which means our system can generate tasks with descriptions much more similar to human-authored tasks compared to the baselines. This suggest our system is effective in extracting tasks that cover ones that human users consider as important and relevant, which is an important requirement for supporting interactive task understanding (discussed in Section~\ref{sec:ux}).
}
\begin{table}[h!]
  \caption{\rev{\textsc{Rouge-L} F-measure and \textsc{Meteor} of our methods and baselines when compared against human-provided tasks in MoTIF test set.
  Experiments that involve randomness are reported in mean$_\text{std}$ format over five runs.}}
  \Description{This is a table that compares the \textsc{Rouge} scores of our system compared to baselines on the MoTIF dataset’s test set. Our system achieves 0.468 \textsc{Rouge-L} score, which are significantly higher than the baselines of random-trace at 0.250  and element-text at 0.066.
  There is a similar observation on the \textsc{Meteor} metric as well.}
  \label{tab:rouge}
  \begin{tabular}{rll}
    \toprule
    Method & \textsc{Rouge-L} & \textsc{Meteor}\\
    \midrule
    \textbf{Ours} & $\textbf{0.468}_{0.005}$ & $\textbf{0.379}_{0.003}$ \\ 
    Random-Trace & $0.250_{0.010}$ & $0.151_{0.007}$ \\
    Element-Text & 0.066 & 0.155 \\
  \bottomrule
\end{tabular}
\end{table}
\subsection{Qualitative Examples}
\label{sec:qual}
\begin{figure*}[h!]
  \centering
  \includegraphics[width=0.87\linewidth]{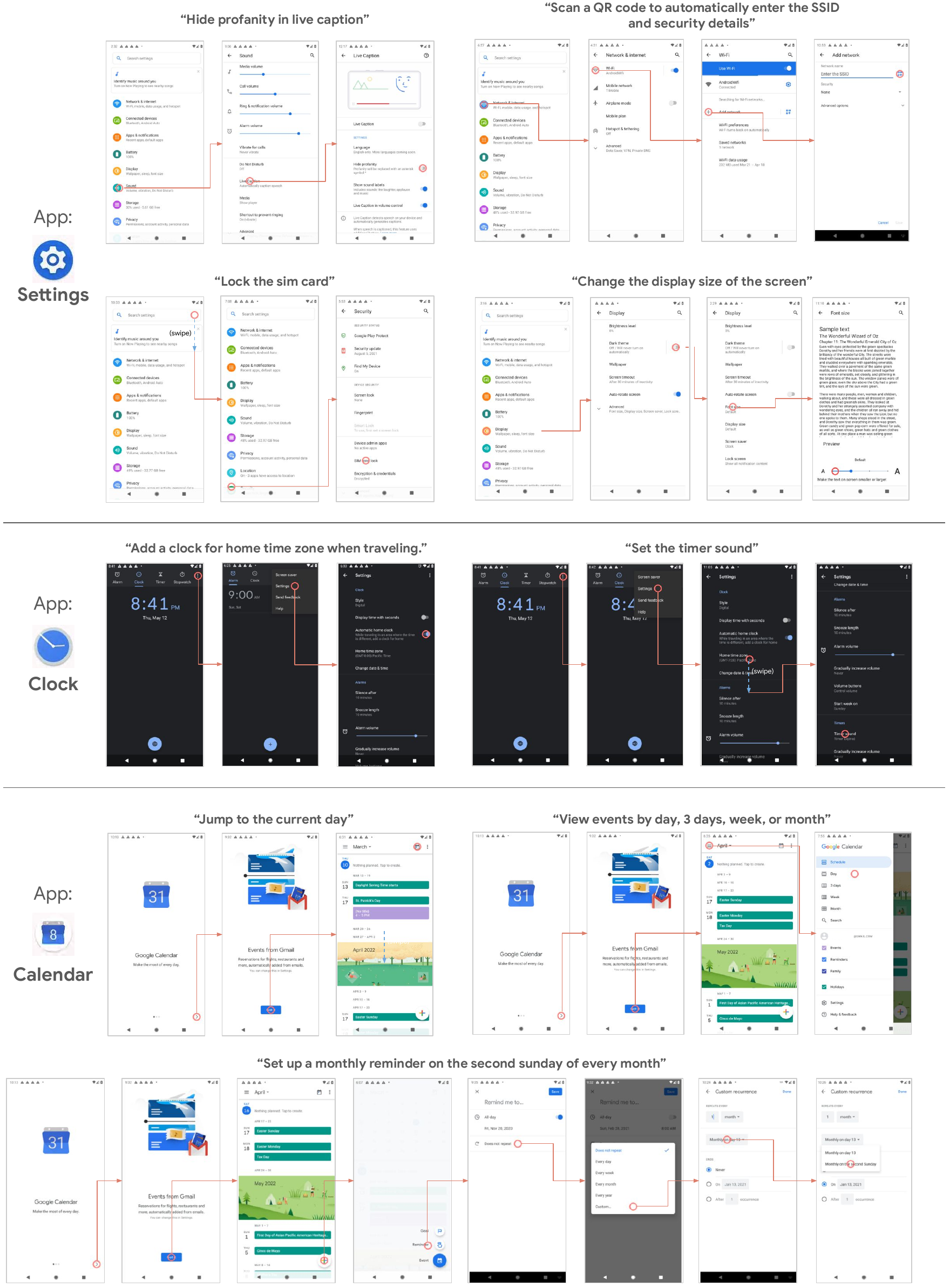}
  \caption{Macro extraction results for apps in the Rehearsal dataset.}
  \Description{This is a large figure with examples of Macros extracted by our system from three apps in the Rehearsal dataset. At the top are macros from the settings app, which are screenshots of the settings page in the Android OS. The four tasks shown are ‘hide profanity in live caption’, ‘Scan a QR code to automatically enter the SSID and security details”, “lock the sim card” , and “Change the display size of the screen”. In the middle are tasks from the Android clock app, which are respectively ‘add a clock for home time zone when traveling” and “set the timer sound”. In the bottom are macros from the calendar app, which are respectively “jump to the current day”, “view events by day, 3 days, week, or month”. In the bottom of the entire figure is a complex trace from the calendar app for ‘Set up a monthly reminder on the second sunday of every month”, which includes eight screens with multiple steps to configure this task in the calendar app.}
  \label{fig:rehearsal}
\end{figure*}

\begin{figure*}[h!]
  \centering
  \includegraphics[width=0.9\linewidth]{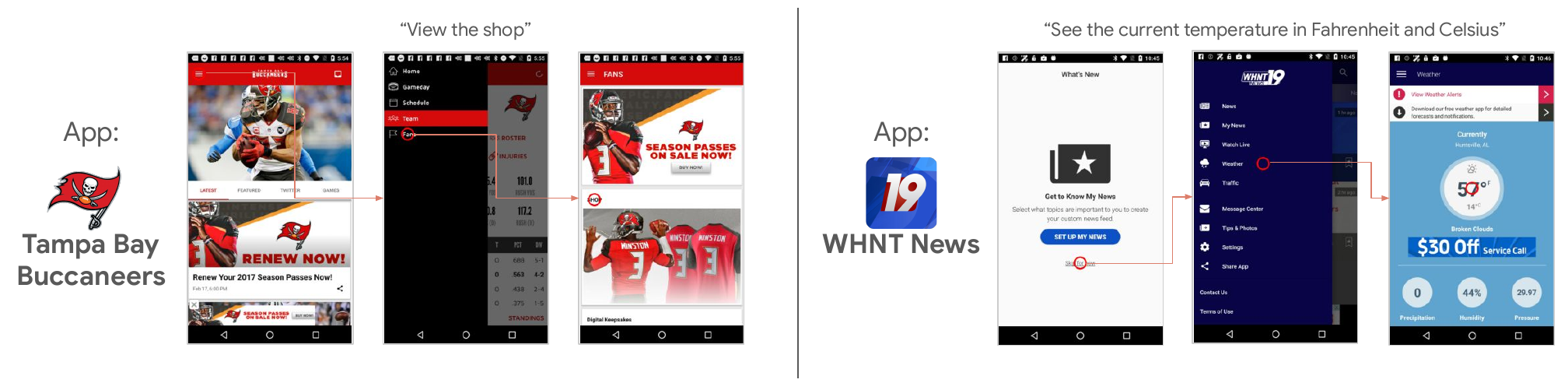}
  \caption{Macro extraction results for apps in the RICO dataset.}
  \Description{This is a figure with examples of macros extracted by our system from two apps in the RICO dataset. On the left is a macro ‘view the shop’ extracted from the Tampa Bay Buccaneers  app, which is an app of a football team and involves three steps in the task, with screenshots shown in the figure to first open the menu on the top right, then click on various buttons. On the right is a macro ‘see the current temperature in Fahrenheit and Celsius’ for the WHNT news app, which includes three screenshots of our extracted actions of navigating to the weather screen and seeing the temperature.}
  \label{fig:rico}
\end{figure*}

To further demonstrate the effectiveness of our extracted macros and show the validity of the macro actions, we qualitatively evaluate multiple macros extracted from the Rehearsal dataset (Figure~\ref{fig:rehearsal}) and the RICO dataset (Figure~\ref{fig:rico}) respectively. In Figure~\ref{fig:rehearsal}, we observe that our system is able to extract highly complex and non-trivial, hidden macros that are useful for the user---in the settings app, our system is able to extract multiple traces that have features unobvious to the user, such as using a QR code to enter SSID and security details, hiding profanity in the live caption feature, and changing the display size of the screen. This observation is further echoed by our findings from the user study, such that a significant proportion of the tasks were not known to the participants prior to the study in Section~\ref{sec:study}. This provides evidence for our system and macro dataset to support applications in how-to knowledge sharing (discussed in Section~\ref{sec:howto}).

In the calendar app, it is able to extract some highly complex macros with a large number of steps, such as setting a monthly reminder on the second sunday of every month. In addition, our extraction system is also able to properly handle scrolling, in which it is able to scroll down through the settings to expose the appropriate elements to `set a timer sound'. These are important features for supporting UI automation (discussed in Section~\ref{sec:automation}). 

In Figure~\ref{fig:rico}, we further show our system's applicability to a wider range of apps with varying features from the RICO dataset. In the app built for the American Football team \emph{Tampa Bay Buccaneers}, our system was able to extract a multi-step macro that opens the shop of the team. In the \emph{WHNT news} app, the system is able to infer the app's feature to display temperature in either the Celsius or Fahrenheit scale by simply observing the temperature reading texts from the UI. These macros demonstrate the ability of our system to understand common user tasks and mobile app UIs, and make appropriate predictions for relevant macros and tasks in broader contexts.

\rev{
\subsection{Trace Optimality}
An important feature of our system is to extract optimal paths for executing the macro tasks by building interaction graphs. While there are no ground-truth data for optimal paths currently available for the tasks, we report the statistics of our macros in RICO and Rehearsal dataset before (pre-optimized) and after (post-optimized) running the Action/Trace merging steps introduced in Section~\ref{sec:optim}. The extracted macros from the RICO dataset contain 6.05 actions pre-optimization on average, and they contain 3.41 actions post-optimization. This represents a 43.6\% reduction in the final macro lengths. The extracted macros from the Rehearsal dataset contain 7.51 actions pre-optimization on average, and they contain 3.40 actions post-optimization. This represents a 54.7\% reduction in the final macro lengths. Our system is able to effectively reduce the required effort for performing the macro tasks by removing around half of the steps in the macros from both datasets.
}


\subsection{User Evaluation of Extracted Macros}
\label{sec:study}
To more comprehensively validate the effectiveness of our system in extracting useful macros, we conducted a user study to gather users' ratings on the quality of our macros. We randomly sampled 60 extracted macros from 3 apps in the Rehearsal dataset. We chose apps from the Rehearsal dataset since they are mostly system apps that are well-known to most mobile users. The scale of macros extracted from these apps also allows us to evaluate each of them in depth. \rev{We also conducted the same study on the PixelHelp dataset that contains descriptions and human-curated tasks, which can be considered as the `golden standard' of the target macros that our system aims to extract; we comparatively analyze the results in this section.
}

For each interaction trace/macro, we show our participants the optimized/human-curated trace screenshots and the task description, and ask them to provide answers for the following questions:

\begin{itemize}
    \item Q1. On a scale of 1-5, does the task match the trace well?
    \item Q2. On a scale of 1-5, is this a reasonable task and trace for mobile users?
    \item Q3. Did you know how to perform this task before seeing this example? (Yes/No)
\end{itemize}

\rev{We also collected open-ended feedback at the end of the study. We recruited 11 participants to evaluate each set of macros through a mailing list internal to our organization. Our participants have 9-20 years of experience using smartphones ($\mu = 13.0 \text{ years}$). Two of our participants are experts in app testing, hence highly familiar with the tasks on these apps. We advised our participants to take approximately 60 minutes to complete the study. Each participant received compensation equivalent to \$40 USD in value. 

The results are summarized in Table~\ref{tab:study}. Our participants rated \textbf{3.85} out of 5 for the correspondence between the descriptions and the actions in the macros (Q1). \textbf{Over 70\%} of the responses for this question have ratings equal or above $4$ (corresponds to the description), demonstrating our system's ability to extract valid tasks that matches the macros' descriptions.

Our participants rated \textbf{4.11} out of 5 for the reasonableness of the macros (Q2) on average, with \textbf{over 75\%} of the responses having ratings equal or above $4$, and \textbf{over 50\%} having ratings of $5$ (considered highly reasonable by the users). This shows that our system is able to effectively extract reasonable macros from the apps in general. Moreover, both correspondence and reasonableness ratings (Q1 and Q2) attained by the macro extracted by our proposed system approach those obtained by tasks in the PixelHelp dataset, reflecting a similar quality of extracted Macros compared to human-extracted tasks in the PixelHelp dataset~\cite{seq2act}. Nevertheless, participants reflect that `some of the clicks (in the traces) seemed inefficient, or (they were) unsure why the UI would click in a certain area' (P1), which leads to the remaining quality gap between human-curated tasks and macros mined automatically by our system. We believe such inefficiency originates from the limitations of our systems, which will be further discussed in Section~\ref{sec:limit}.

Finally, an important and interesting finding from this study is that our participants \emph{did not know} how to perform \textbf{48.2\%} of the tasks prior to seeing the macros (Q3). This percentage is also comparable to that of the PixelHelp dataset. This reveals a great potential for our system to support task discovery and tutorial generation (discussed in Section~\ref{sec:howto}), such that almost half the tasks discovered by our system are non-trivial tasks supported by the app, similar to the collection of human-curated tasks (PixelHelp). Our system can automatically find tasks that users are unfamiliar with and present reasonable steps for them to complete the tasks, all coupled with conversational interactions given the high-level descriptions also extracted by our system.

\begin{table*}[h!]
  \caption{User study results (mean$_\text{std}$) for the PixelHelp Dataset (human-curated tasks)~\cite{seq2act} and our extracted macros.}
  \Description{This is a table that summarizes the ratings of user study's first two questions and the percentage of unknown tasks seen by the participants for PixelHelp macros and our extracted macros.}
  \label{tab:study}
  \begin{tabular}{rlll}
    \toprule
   \textbf{Macro Source} & \textbf{Q1. Correspondence (?/5)} $\uparrow$ & \textbf{Q2. Reasonableness (?/5)} $\uparrow$  & \textbf{Q3. Unknown Task \%} \\
    \midrule
    PixelHelp (Human-Curated) & 4.12$_{1.23}$ & 4.25$_{1.10}$ & 47.4\%$_{49.9\%}$ \\
    \midrule
    \textbf{Ours} & 3.85$_{1.40}$ & 4.11$_{1.15}$ & 48.2\%$_{50.0\%}$ \\ 
  \bottomrule
\end{tabular}
\end{table*}
}

\subsection{Execution Success}
While extracting representative and relevant macro descriptions is an important first step towards mining useful macros, the end-goal of having these macros is to be able to execute them in real environments. Replayable macros provide the possibility for models and systems to operate within the natural language modality and interact with mobile apps and devices effectively, supporting automation for interactions (discussed in Section~\ref{sec:automation}).

Towards this goal, we built a live environment to examine the success rate of executing the extracted macro in Android devices, hence evaluating the end-to-end performance of our system. We took 60 extracted macros from the Rehearsal dataset, re-executing them in the live environment, and recording the success rate for the Macro Replayer in reaching the final screen of the macros by manually inspecting them. The live environment was developed and deployed as Android emulators, using Pixel 4 devices with Android version 30.

On average, 76.7\% of the macros extracted by our system were successfully executable in the live environment. This not only shows that a large proportion of extracted macros are valid and are replayable in realistic environments, but also the high effectiveness of our Macro Replayer in automating the tasks specified in the macros.

\section{Envisioned Applications}
\label{sec:application}
Based on our system development and findings, we envision multiple future applications supported and enabled by our contributions.

\rev{\subsection{Interactive Task Understanding and Modeling}
\label{sec:ux}
Our contributed dataset expands UI understanding of existing datasets to \emph{interactive task understanding} that involve text descriptions and multiple screens/actions. We envision this dataset to be capable of supporting multiple downstream machine-learning applications, including the training of agent models that can act on UIs based on high-level goals for navigation and task completion. Other than agent models, we also envision the dataset to support generative design applications. While existing works focus on generating individual UI designs, annotated traces with semantically meaningful tasks can be used to train models that can generate \emph{sequences} of different UI screens that supports high-level tasks and/or user-stories, greatly expanding the applicability of these models in realistic design scenarios. 
}

Moreover, we also envision our contributions to help facilitate and accelerate UX development processes. For example, UX designers can use the proposed method to visualize and test macros on app prototypes, or search for similar macros in other apps.

\subsection{Interaction Automation}
\label{sec:automation}
Another major area that we envision our work to support is the automation of user interactions, such as enabling conversational task-based interactions and task shortcuts for apps. While conversational interactions can currently be effectively supported by LLMs, acting directly in the UI space is challenging for LLMs. Our methods enable task-based interactions in the natural language space, such that LLMs can directly select and rank macros by their text descriptions based on existing conversations with users (which is a space more familiar to LLMs), and the macros can be directly executed on apps with potentially higher success rate compared to having the LLMs directly act on the UI elements.

Task shortcuts are another important area of UI automation explored by prior works~\cite{savant}. Nevertheless, to our knowledge none of them allow for the automatic discovery of new tasks and they require users to provide task descriptions to utilize the shortcuts. Our methods can enable a new set of applications that do not assume users to possess comprehensive prior knowledge of the tasks.

\subsection{How-to Knowledge Sharing}
\label{sec:howto}
\rev{Beyond understanding interactive tasks and supporting automation, our proposed method can support the discovery and sharing of how-to knowledge in apps, such as automatically generating tutorials in apps without requiring additional effort of the app developers. Our approach can be coupled with automatic crawlers to obtain UIs without users' manual exploration (such as the approach we took for the Rehearsal dataset in Section~\ref{sec:dataset}), and hence can discover macros that correspond to tasks unseen by users. Our findings in Section~\ref{sec:study} also validate that a large proportion of tasks fully-automatically discovered by our system were novel to the users, and we envision our system to be used to educate users in performing new tasks and/or taking more optimal paths in known tasks.}



\section{Limitations and Future Work}
\label{sec:limit}
\begin{figure*}[h!]
  \centering
  \includegraphics[width=0.8\linewidth]{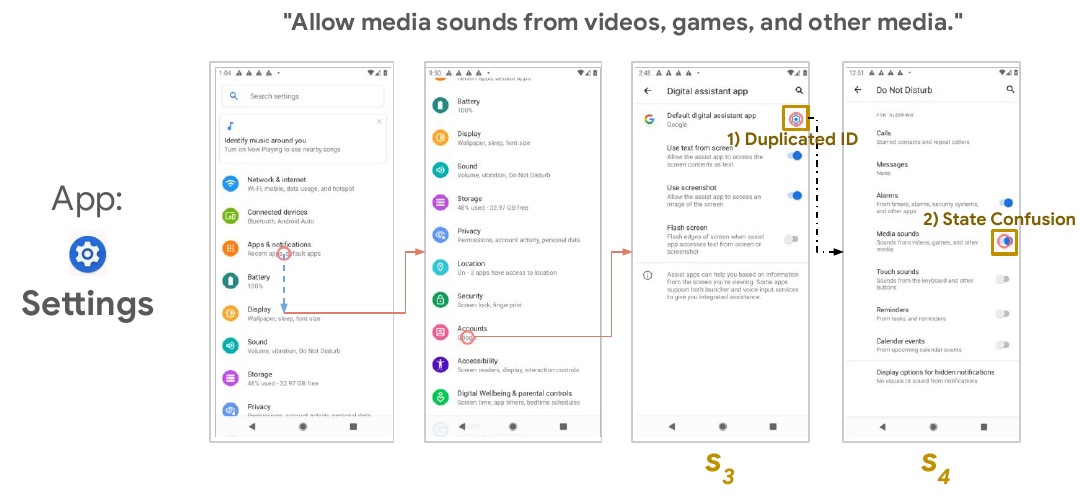}
  \caption{\rev{Limitations of our current macro mining system illustrated by a single Macro example.}}
  \Description{This figure shows a flawed extraced macros with limitations of our system. The macro displayed is from the settings app with the description `allow media sounds from videos, games, and other media.' In the third screen, there is a settings icon that is on the digital assistant app. However, clicking on this incorrectly leads to the fourth screen of do not disturb. This is a limitation from UI elements having duplicated IDs in the macros. In addition, in the last screen s4, clicking on a switch that is enabled is suggested for the macro, which is incorrect since the macro is to `allow' media sounds, while clicking on a switch already enabled disables it.}
  \label{fig:limitation}
\end{figure*}

Our work has several limitations that warrant future investigation. The first limitation is the limited representational power and uniqueness of the element/node identifiers. We currently rely on resource\_ids and nearby text contents through a carefully designed algorithm to identify identical UI actions, but even this can be insufficient at times when resource\_id and text content are non-unique, resulting in the incorrect merging and path-finding of some macros that pass through these nodes with duplicated resource\_ids that should have been unique. \rev{For example, in the macro in Figure~\ref{fig:limitation}, the action performed on the third screen $s_3$ was incorrect, such that clicking on the `settings' button (marked with 1) in the Figure) does not lead to the next screen $s_4$. This is because the fourth screen $s_4$ can be reached from another `settings' button that looks identical and with an identical resource\_id, but was used in a different screen and different context. This led to an invalid optimization performed by our system. Future work shall explore the inclusion of dynamic amount of context around the elements, or utilize learning-based approaches (e.g., element embeddings~\cite{widget-captioning, seq2act, screen2vec} with thresholding) to identify similar UI elements for merging.

Another important limitation of our system is the assumption that all actions are stateless and will lead to the same result when the same actions are performed on it. This assumption might not be true for some apps, such as those that contain dynamic content. For instance, the macro in Figure~\ref{fig:limitation} is to `allow' the `media sounds' setting. However, clicking on the switch on the fourth screen $s_4$ will lead to the system disabling media sounds since the setting is already enabled. Adequately representing and handling context information of the actions is another important area of future work.
} 

\section{Conclusions}
In this paper, we introduced a novel LLM-based system for automatically extracting semantically meaningful macros from any interaction traces. Our system first extracts meaningful tasks from each UI in the app, and consequently merges them into a compact and comprehensive set of efficient macros. Our evaluation shows that the extracted macros are highly relevant to realistic human-curated tasks when comparing against an existing large-scale dataset. The majority of the macros can also be successfully executed, and are considered to be reasonable and valid by the participants of our user study. \rev{We believe the dataset and the system can support important future applications in interaction automation, interactive task understanding, and how-to knowledge sharing.}

\begin{acks}
We thank all anonymous reviewers for their constructive comments and suggestions in the paper review process. We would also like to thank Chin-Yi Cheng for his suggestions and assistance in this project, and all participants in our user study for evaluating macros and providing feedback.
\end{acks}

\bibliographystyle{ACM-Reference-Format}
\bibliography{samples/sample-base}

\appendix
\rev{
\section{Macro Filtering} \label{appx:macro_filtering}
Here, we describe the heuristic used to filter macros in Sec.~\ref{sec:task_extraction}. In general, we want to retain macros that are specific and filter out those only contain general words. To do so, we first define a list of common phrases:

\egboxwide{
  setting, settings, app, apps, element, elements, text, input, field, button, image, screen, left, right, top, bottom, top-left, top-right, bottom-left, bottom-right, previous, next, close, cancel, tap, press, click, confirm, set, enter, navigate.
  }

and a list of non-content words:

\egboxwide{
    is, the, are, and, else, with, to, on, in, at, off, within, without, below, above, up.
}

Firstly, we remove non-content words from a macro.
Then we check if the remaining content words are exclusively all common phrases. If so, we simply remove this macro.
Finally, we check if the underlying macro has any action that matches any of the following keywords:

\egboxwide{
    cancel, go back, back, go\_back, prev, previous, navigate up, navigate\_up, try again, (id=.
}

Doing so helps up avoid cycling trajectories and erroneous ones from LLM inference.

\section{Low \textsc{Rouge-L} score examples}
\label{appx:rouge}
We show three examples with some of the lowest \textsc{Rouge-L} score attained by our extracted macro, when compared against the MoTIF dataset in Section~\ref{sec:quant} in Figure~\ref{fig:rouge}. In some of these examples, we observe that some of tasks extracted by our model is reasonable, but was only missing details required in the MoTIF dataset, such as the flight destination and date detail required in example a), and the specific hotel detail in example b). Moreover, some MoTIF traces are incomplete, such as example c), and based on the incomplete traces, we believe our system's extraction of `review the app's privacy policy' is perhaps more reasonable than the ground-truth, since there is neither temperature forecast nor the location of New Orleans shown in example c).

\begin{figure*}[t!]
  \centering
  \includegraphics[width=0.87\linewidth]{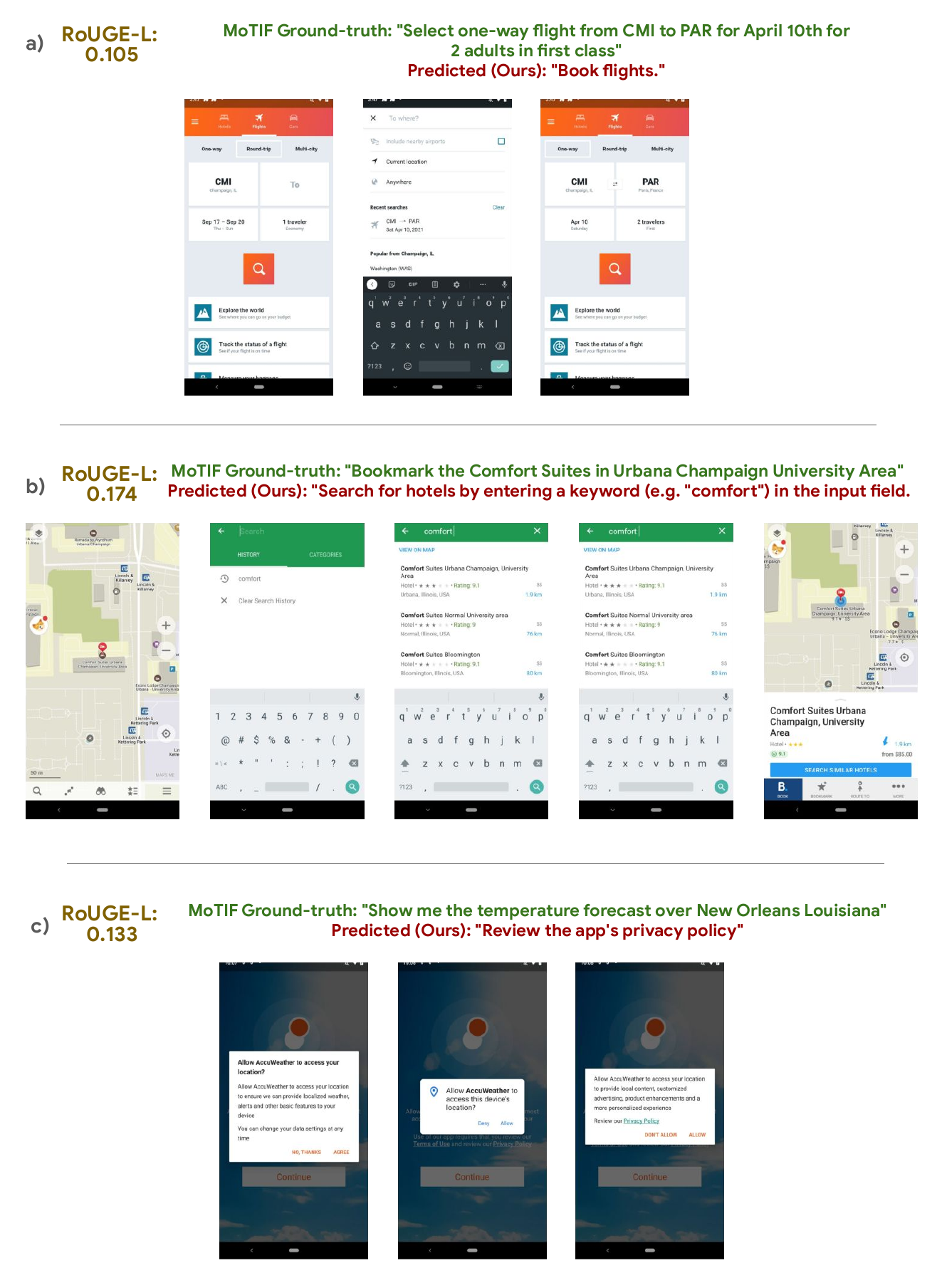}
  \caption{Examples with low \textsc{Rouge-L} scores.}
  \Description{This figure inclues three traces with low Rouge-L scores. The top trace a is a trace for booking flights, with grid UIs of a search page as the first UI, and a list UI showing results in the second page. The middle trace b is a UI for searching for hotels, with keyboards and search bars in the UI and a list of results of `Comfort Suites'. The bottom trace c is a trace that has pop up dialogs asking users for various permissions to support the app. All of these traces have low rouge scores, yet the predicted macros are still reasonable.}
  \label{fig:rouge}
\end{figure*}

\begin{table*}[h!]
\caption{LLM inputs and outputs example for the `create a reminder' example macro.}
\Description{This table shows LLM inputs and outputs from the task discovery and action grounding steps in the macro mining process.}
\label{tab:llm1}
\centering
\begin{tabularx}{\textwidth}{L{0.1}|L{0.65}|L{0.25}}
 \toprule
 \textbf{Substep} & \textbf{LLM Input} & \textbf{LLM \newline Output} \\ \midrule
3.2.1 Task Discovery & Below is a simplified HTML code of a mobile app: \newline
<screen> \newline
<img id="0" class="cancel image" pos="top left"></img> \newline
<button id="1" class="save" pos="top right">save</button> \newline
<input id="2" class="title edit" pos="top">remind me to</input> \newline
<p id="3" class="first line" pos="top">all day</p> \newline
<img id="4" class="tile icon" pos="top left"></img> \newline
<p id="5" class="first line" pos="top">sun dec 13 2020</p> \newline
<button id="6" alt="start time 8 00 am" pos="top right">8 00 am</button> \newline
<p id="7" class="first line" alt="does not repeat" pos="top">does not repeat</p> \newline
<img id="8" class="tile icon" pos="top left"></img> \newline
</screen> \newline What can a user do with the prompt? \newline The user can: -  & create a reminder \newline
- edit the reminder title \newline
- set the reminder time \newline
- set the reminder date \newline
- choose whether the reminder repeats \\ \midrule
3.2.2 Action Grounding & Below is a simplified HTML code of a mobile app: \newline
<screen> \newline
<img id="0" class="cancel image" pos="top left"></img> \newline
<button id="1" class="save" pos="top right">save</button> \newline
<input id="2" class="title edit" pos="top">remind me to</input> \newline
<p id="3" class="first line" pos="top">all day</p> \newline
<img id="4" class="tile icon" pos="top left"></img> \newline
<p id="5" class="first line" pos="top">sun dec 13 2020</p> \newline
<button id="6" alt="start time 8 00 am" pos="top right">8 00 am</button> \newline
<p id="7" class="first line" alt="does not repeat" pos="top">does not repeat</p> \newline
<img id="8" class="tile icon" pos="top left"></img> \newline
</screen> \newline
Which element id(s) should the user click on next to accomplish the task Create a reminder? \newline
Respond with only the number(s), or "None" if the user can already complete the task on the current page. & 1 \\ \midrule \end{tabularx}
\end{table*}

\section{Full LLM Inputs and Outputs Example} \label{appx:llm}
Tables~\ref{tab:llm1} and~\ref{tab:llm2} lists the full LLM inputs and outputs for the extraction system on $s_9$ of Figure~\ref{fig:pipeline}, the representative macro example of `create a reminder' used in Section~\ref{sec:pipeline}.

\begin{table*}[bh!]
\caption{LLM inputs and outputs example for the `create a reminder' example macro (continued).}
\Description{This table continues from the previous table to show LLM inputs and outputs from the parameter extraction step in the macro mining process.}
\label{tab:llm2}
\centering
\begin{tabularx}{\textwidth}{L{0.1}|L{0.65}|L{0.25}}
 \toprule
 \textbf{Substep} & \textbf{LLM Input} & \textbf{LLM \newline Output} \\ \midrule
3.2.3a \newline Parameter Text & Below is a simplified HTML code of a mobile app: \newline
<screen> \newline
<img id="0" class="cancel image" pos="top left"></img> \newline
<button id="1" class="save" pos="top right">save</button> \newline
<input id="2" class="title edit" pos="top">remind me to</input> \newline
<p id="3" class="first line" pos="top">all day</p> \newline
<img id="4" class="tile icon" pos="top left"></img> \newline
<p id="5" class="first line" pos="top">sun dec 13 2020</p> \newline
<button id="6" alt="start time 8 00 am" pos="top right">8 00 am</button> \newline
<p id="7" class="first line" alt="does not repeat" pos="top">does not repeat</p> \newline
<img id="8" class="tile icon" pos="top left"></img> \newline
</screen> \newline The user is trying to complete the task create a reminder. \newline Other than clicking on the element with id 1, list the additional information the user needs to enter in the format of (- (info)). Answer "None" if no additional information is needed. & - (title) \newline - (date) \\ \midrule
3.2.3b \newline Parameter \newline Element & Below is a simplified HTML code of a mobile app: \newline
<screen> \newline
<img id="0" class="cancel image" pos="top left"></img> \newline
<button id="1" class="save" pos="top right">save</button> \newline
<input id="2" class="title edit" pos="top">remind me to</input> \newline
<p id="3" class="first line" pos="top">all day</p> \newline
<img id="4" class="tile icon" pos="top left"></img> \newline
<p id="5" class="first line" pos="top">sun dec 13 2020</p> \newline
<button id="6" alt="start time 8 00 am" pos="top right">8 00 am</button> \newline
<p id="7" class="first line" alt="does not repeat" pos="top">does not repeat</p> \newline
<img id="8" class="tile icon" pos="top left"></img> \newline
</screen> \newline Where can the user enter the title? Answer with only the element id, or "None" if no element matches. & 2 \\ \midrule
\end{tabularx}
\end{table*}

}
\end{document}